# Thermal Conductivity of BAs under Pressure

*Songrui Hou, Bo Sun, Fei Tian, Qingan Cai, Youming Xu, Shanmin Wang, Xi Chen, Zhifeng Ren, Chen Li*, Richard B. Wilson**

S. Hou, Prof. C. Li, Prof. R. B. Wilson

Materials Science and Engineering, University of California, Riverside, California 92521, USA

Email: chenli@ucr.edu and rwilson@ucr.edu

Prof. B. Sun

Tsinghua Shenzhen International Graduate School, Tsinghua University, Shenzhen, Guangdong 518055, China

Prof. F. Tian

School of Materials Science and Engineering, Sun Yat-sen University, Guangzhou, Guangdong 510275, China

Prof. F. Tian, Prof. Z. Ren

Department of Physics and Texas Center for Superconductivity, University of Houston, Houston, Texas 77204, USA

Q. Cai, Prof. C. Li, Prof. R. B. Wilson

Department of Mechanical Engineering, University of California, Riverside, California 92521, USA

Y. Xu, Prof. X. Chen,

Department of Electrical and Computer Engineering, University of California, Riverside, California 92521, USA

Prof. S. Wang

Department of Physics, Southern University of Science and Technology, Shenzhen, Guangdong 518055, China

**Abstract**

The thermal conductivity of boron arsenide (BAs) is believed to be influenced by phonon scattering selection rules due to its special phonon dispersion. Compression of BAs leads to significant changes in phonon dispersion, which allows for a test of first principles theories for how phonon dispersion affects three- and four-phonon scattering rates. This study reports the thermal conductivity of BAs from 0 to 30 GPa. Thermal conductivity vs. pressure of BAs is measured by time-domain thermoreflectance with a diamond anvil cell. In stark contrast to what is typical for nonmetallic crystals, BAs is observed to have a pressure independent thermal conductivity below 30 GPa. The thermal conductivity of nonmetallic crystals typically increases upon compression. The unusual pressure independence of BAs's thermal conductivity shows the important relationship between phonon dispersion properties and three- and four-phonon scattering rates.

**Introduction**

High-thermal-conductivity materials are desirable for thermal management applications. Power electronic devices operate at power densities higher than 100 W/cm$^2$, roughly three orders of magnitude larger than the irradiance of the Sun[1,2]. Discovery and integration of high thermal conductivity materials into electronics offer a route for increasing performance. Discovery of such materials requires a detailed understanding of material properties that lead to high thermal conductivity. However, despite more than a half century of study, a complete microscopic understanding does not exist for why some materials have high thermal conductivity, while other similar materials do not. Our study aims to help fill this fundamental gap by experimentally testing the relationship between BAs's phonon dispersion and phonon scattering rates.

In the 1970s, Slack came up with four rules for finding nonmetallic crystals with high thermal conductivity. These rules are: i) low average atomic mass, ii) strong interatomic bonding, iii) simple crystal structure, and iv) low anharmonicity[3]. Given their simplicity, the apparent accuracy of Slack's rules has been something of a long-standing puzzle. Theoretical models for phonon-phonon scattering rates have long predicted that, because phonon scattering processes must conserve energy and crystal momentum, phonon dispersion can have a strong effect on phonon scattering rates[4,5]. Certain features in the phonon dispersion can make it difficult for a three-phonon scattering processes to satisfy selection rules[4,5]. It is well known that crystals with similar crystal structures and average atomic mass can have distinct differences in phonon dispersion[6]. But Slack's rules imply that differences in phonon dispersion between such crystals will have little effect on thermal transport.

In the past ten years, a number of theoretical and experimental studies have started to unravel this puzzle and correct Slack's rules[7–16]. In 2013, Lindsay *et al.* used first principles calculations based on density functional theory (DFT) and the Peierls-Boltzmann equation (PBE) to study the effects of atypical phonon dispersion on thermal transport[7–9]. Their first-principles based work predict that phonon dispersion relations have a strong effect on scattering rates via selection rules[7–9]. For example, first principles theory predicts that crystals with special phonon dispersion properties like BAs will have a thermal conductivity higher than the value Slack's rules predict[8,17]. BAs's phonon dispersion is special for two reasons. First, BAs has a large frequency gap between acoustic and optic phonons. This gap should eliminate the phase space for three-phonon scattering between acoustic and optic phonons[8,18]. Second, the acoustic phonon branches of BAs are unusually close together[18]. This acoustic bunching effect is predicted to result in a small phase space for three-phonon scattering processes of acoustic phonons[8,19].

Several recent experimental studies have verified first principles predictions that BAs has an anomalously large thermal conductivity[12–14]. BAs has a thermal conductivity between 1000 and 1300 W m$^{-1}$ K$^{-1}$, see Figure 1. Despite similar average atomic mass, bonding, and crystal structure, BAs has a thermal conductivity ~7× larger than Silicon.

The good agreement between experiment and first-principles theory for the thermal conductivity of BAs, and a number of other high thermal conductivity materials[9,12–16], provides compelling indirect evidence that a strong relationship exists between phonon dispersion properties and thermal conductivity. However, so far, there are no experimental studies that directly test the hypothesis that acoustic bunching leads to higher thermal conductivity. Testing of this hypothesis requires systematically tuning a material's phonon dispersion relation and observing the subsequent changes in thermal conductivity.

The pressure dependence of BAs's thermal conductivity offers a way to experimentally explore the relationship between phonon dispersion, phonon scattering selection rules, and thermal transport. Thermal conductivity is a weighted average of phonon lifetimes. Therefore, measurements of thermal conductivity vs. pressure provides some indirect information about how phonon lifetimes depend on pressure. First principles calculations show that pressure stiffens longitudinal acoustic phonons, thereby increasing the energy difference between longitudinal and transverse acoustic phonons[10,11]. In other words, compression of BAs reduces acoustic bunching, see Figure 1(a), and makes BAs's dispersion relation more like a typical crystal, *e.g.*, Si. Furthermore, pressure dependent measurements of BAs's thermal conductivity also offer the opportunity to study how phonon dispersion affects four-phonon scattering processes. Four phonon scattering processes are believed to play an important role in BAs[12–14,17]. Four phonon scattering rates are believed to depend on phonon dispersion properties such as the frequency gap between acoustic and optic modes[9,10].

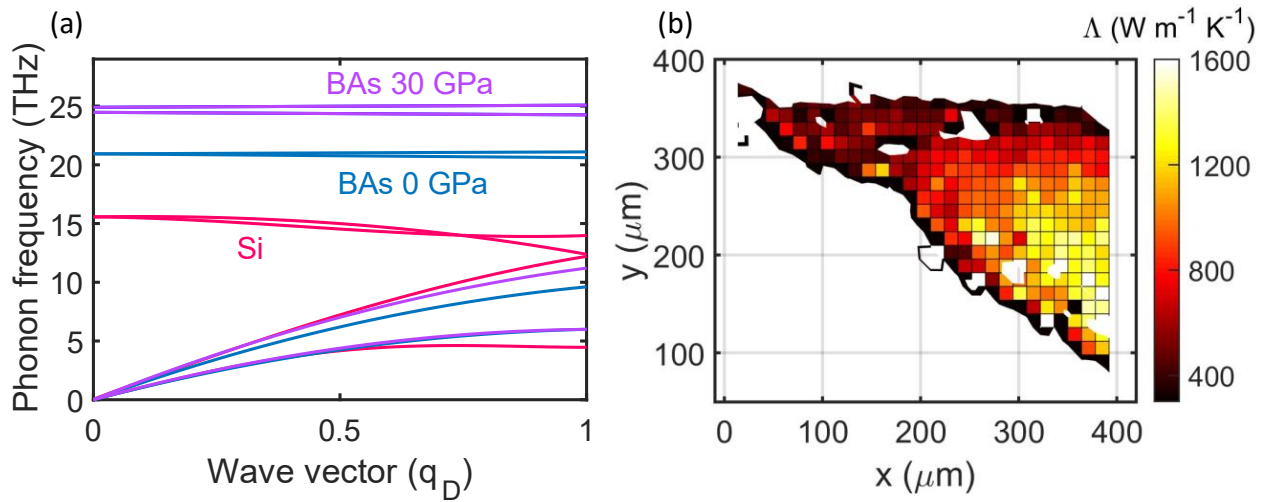

Figure 1. (a) Approximate phonon dispersion relations of BAs at 0 and 30 GPa. $q_D$ refers to the wave vector at the zone boundary. We construct isotropic dispersions of BAs by imitating the DFT calculation results of BAs in Ref. [10]. Stiffening of the longitudinal acoustic phonon branch with increasing pressure reduces acoustic bunching. The dispersion relations of Si at 0 GPa are included as red curves[20]. The large *a-o* gap and acoustic bunching are believed to lead to the high $\Lambda$ of BAs. Increased pressure reduces the bunching of BAs's acoustic modes, leading to a dispersion relation more like Si. (b) Map of the thermal conductivity at 0 GPa of the triangular BAs crystal that is the focus of our study. The thermal conductivity in the interior of the crystal ranges between 1000 and 1300 W m$^{-1}$ K$^{-1}$.

The aim of our experimental study is to investigate phonon scattering mechanisms in BAs using high pressure. We perform time-domain thermoreflectance (TDTR) measurements of BAs in a diamond anvil cell (DAC) (Figure 2). TDTR is a well-established tool for measuring thermal conductivity[21]. Diamond anvil cells can generate pressures on the scale of GPa. We include detailed descriptions of our experiments in Methods. We report the thermal conductivity as a function of pressure, $\Lambda(P)$, for three BAs samples with different ambient thermal conductivities. The apparent thermal conductivities of our three samples derived from TDTR measurements at ambient conditions are ~1100, 600 and 350 W m$^{-1}$ K$^{-1}$. We also measure the thermal conductivity of two MgO single crystals as control experiments. The thermal conductivities of all three BAs samples depend weakly on pressure between 0 and 30 GPa. Alternatively, for MgO, we observe a monotonically increasing thermal conductivity with increasing pressure.

The weak pressure dependence of $\Lambda(P)$ for BAs implies that phonon scattering rates have a weak pressure dependence. We credit the weak pressure dependence of phonon scattering rates to how pressure affects three-phonon vs. four-phonon scattering rates. Decreases in acoustic bunching increase three phonon scattering rates. An increase in the frequency gap between acoustic and optic phonons decreases four-phonon scattering rates. The net effect leads to phonon scattering rates to be pressure independent.

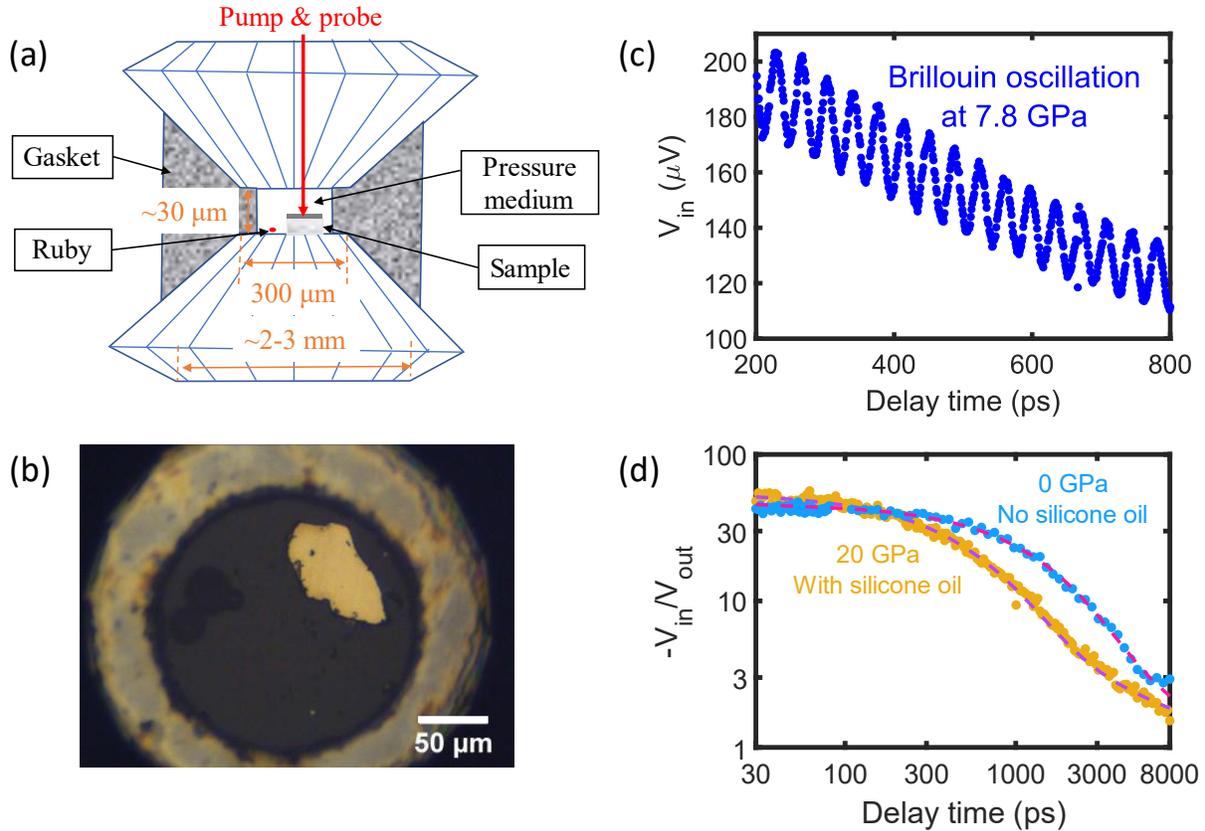

Figure 2. (a) A schematic of the DAC-assisted TDTR measurement. The pump and probe beams transmit through the diamond and pressure medium, and are focused onto the sample surface. (b) Image of a BAs sample coated with Al inside a DAC. We load ruby spheres as pressure indicators. We use a stainless-steel gasket and silicone-oil pressure medium. (c) An example of Brillouin oscillations from the silicone oil in our experimental signal. The frequency of the Brillouin oscillation provides a sensitive measure of local pressure at the sample. (d) TDTR data collected on Sample A at 0 and 20 GPa. The dots and dash lines are the experimental results and the predictions by the heat diffusion model, respectively.

**Results**

The focus of our study is on a high-purity BAs single crystal with thermal conductivity ranging 1000-1300 W m$^{-1}$ K$^{-1}$. We show a thermal conductivity map of this triangular BAs single crystal in Figure 1(b). After collecting the map, we broke the crystal into small pieces and processed one of them for diamond anvil cell (DAC) measurements (Sample A). A schematic of the DAC apparatus and some example TDTR data are shown in Figure 2. Additional details concerning sample preparation are in Methods.

The thermal conductivity of BAs is known to be sensitive to even small concentrations of defects[22–24]. Therefore, to explore how defects affect the pressure dependent thermal conductivity of BAs, we also studied two other crystals with lower ambient thermal conductivities (Samples B and C). The ambient thermal conductivities of these samples are ~600 and 350 W m$^{-1}$ K$^{-1}$.

We observe that high-purity BAs has a constant thermal conductivity of ~1000 W m$^{-1}$ K$^{-1}$ between 0 and 30 GPa, see Figure 3. This is the main result of our study. We also observe that BAs crystals with low concentrations of defects (Samples B and C) have a pressure independent thermal conductivity.

TDTR is a well-established method whose uncertainty depends on input parameters in the heat diffusion model[25–27]. In our measurements, the uncertainty mostly comes from the thickness of Al ($h_{Al}$), heat capacity of Al ($C_{Al}$), laser spot size ($\omega_0$), and heat capacity of BAs ($C_{BAs}$). Typically, we have a ~5% uncertainty in $h_{Al}C_{Al}$[28], ~5% uncertainty in spot size $\omega_0$. We also estimate an uncertainty of ~3% uncertainty for $C_{BAs}$. These yield a total uncertainty in the derived values for $\Lambda_{BAs}$ of ~15%. The error bars in Figure 3 indicate the uncertainty in $\Lambda_{BAs}$.

As described in Methods, we performed multiple TDTR measurements at various locations on the BAs samples at each pressure. The thermal conductivity reported in Figure 3 is the average value from all measurements for a given sample and pressure. We observed that there could be optical artifacts in DAC-assisted TDTR measurements as shown in the -$V_{in}$/$V_{out}$ maps in Figure S6. Detailed discussion of optical artifacts can be found in Supplementary Notes. The purpose of measuring multiple spots is to avoid the optical artifacts and guarantee the reproducibility of our measurements. As expected for a high-quality homogenous crystal, the variance in thermal conductivity at different sample locations is small at most pressures. The TDTR results at each pressure and location are provided in Supplementary Figure S8 to S10.

The apparent thermal conductivity we derive from TDTR measurements of the BAs crystals depend on the size of the laser beam we use in our experiments, see Supplementary Figure S3. The apparent dependence of the thermal conductivity on laser spot size is an artifact caused by the breakdown of the heat-diffusion equation[29,30]. As a result of this spot-size effect, the pressure dependent thermal conductivity values reported in Figure 3 are ~20% lower than the intrinsic value. Our primary interest is the pressure dependence of thermal conductivity. Because the spot-size artifact should not depend on pressure[31], this small deviation does not affect our conclusions. Detailed discussion on the spot-size effect is included in Supplementary Notes.

The weak pressure dependence we observe for all three BAs crystals is in stark contrast with our observations for MgO, see Figure 4. For MgO, we observe a factor of two increase in the thermal conductivity upon compression to 20 GPa. Our results for MgO agree with prior reports[32].

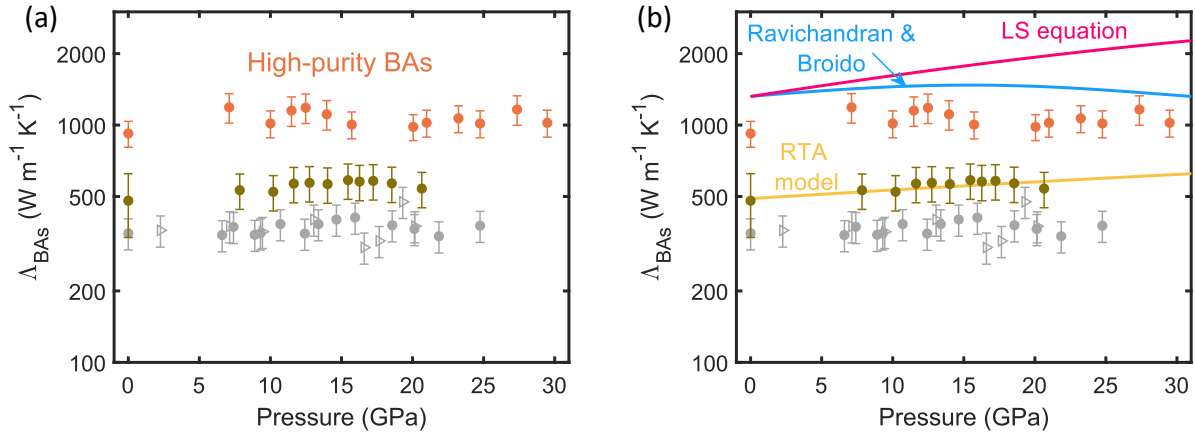

Figure 3. (a) Pressure dependent thermal conductivities of three BAs samples. Dots are compression data while triangles are decompression data. Different colors stand for different samples. (b) Model predictions for the thermal conductivity of BAs. The blue curve is the thermal transport calculation at 300 K from Ref. [10]. The red line represents the Leibfried-Schlömann equation prediction. To show the different trend predicted by first-principles thermal transport calculation and LS equation, we set the ambient Λ values equal. The yellow line is a relaxation time approximation model prediction. The RTA model considers the effects of phonon-phonon scattering and phonon-defect scattering.

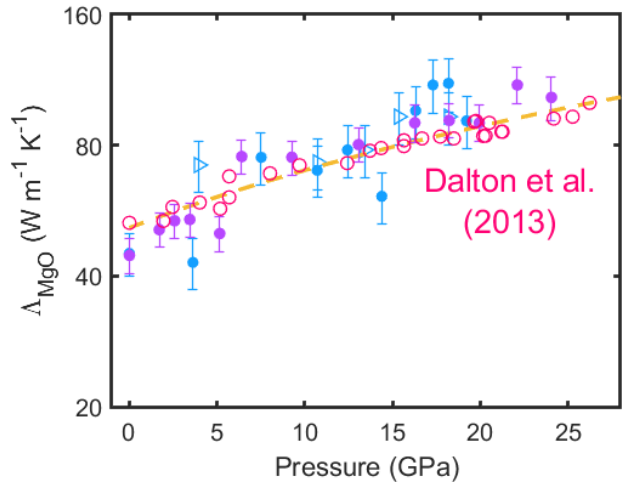

Figure 4. Pressure dependent thermal conductivities of MgO. We characterized two MgO samples as control measurements shown as the blue and purple symbols. Dots and triangles are compression and decompression data, respectively. Here we measured only a single location at each pressure. The error bars here represent the ~10% uncertainties in thermal conductivity that arises from uncertainties in thermal model parameters. Red circles and the orange dashed line are the published experimental data and prediction of the Leibfried-Schlömann equation, both from Ref. [32].

**Discussion**

In the absence of contextualizing information, the dramatic difference in Λ(*P*) for BAs vs. MgO (Figure 3 vs. Figure 4) is quite surprising. The bulk modulus of BAs is 142 GPa, while MgO is 160 GPa[33,34]. Both materials have a relatively small atomic mass per unit cell and simple unit cells. BAs has a zin-blende crystal structure ($F\bar{4}3m$) and MgO has a halite structure ($Fm\bar{3}m$). The Grüneisen parameter of BAs and MgO are both expected to experience a ~10% decrease between 0 and 20 GPa[35,36].

To understand what the difference in Λ(*P*) for BAs vs. MgO implies, it is useful to consider the microscopic origins for a material's thermal conductivity. The thermal conductivity of a material is determined by group velocities, number density, and relaxation times of phonons. Therefore, the pressure dependent thermal conductivity is determined by the pressure dependence of these three vibrational properties. The Leibfried-Schlömann (LS) equation is a simple model for quantifying how these three vibrational properties govern Λ. A number of prior experimental studies show the LS equation often has predictive power in explaining Λ(*P*)[32,37,38]. The LS equation predicts

$$\Lambda = \frac{B\bar{M}\delta\theta^3}{T\gamma^2}. \quad (1)$$

Here *B* is a constant, $\bar{M}$ is the average mass of an atom in the crystal, $\delta^3$ is the average volume occupied by one atom in the crystal, *θ* is the Debye temperature, *T* is temperature, and *γ* is the Grüneisen parameter. We take the pressure dependence of these quantities for BAs from Refs. [36,39]. Not surprisingly, given BAs's special phonon dispersion relation, the LS equation drastically overestimates the Λ(*P*) of BAs, see Figure 3(b). Nevertheless, the LS equation prediction serves as a useful benchmark for what Λ(*P*) should look like if pressure-induced changes to the phonon dispersion do not dramatically alter the phase space for phonon-phonon scattering. We observe that the LS equation does a good job predicting Λ(*P*) in MgO, see Figure 4.

We emphasize that a pressure independent thermal conductivity between 0 and 30 GPa is extremely unusual behavior for nonmetallic materials. Normally, thermal conductivity monotonically increases with increasing pressure[3]. At high pressures, atomic bonds tend to stiffen, and phonon frequencies tend to increase, favoring a higher thermal conductivity. Furthermore, three-phonon scattering rates are governed by phonon anharmonicity. Anharmonicity typically decreases with increasing pressure, as evidenced by the fact that the Grüneisen parameter of most materials tend to decrease upon compression[40]. A reduction in anharmonicity also favors a larger thermal conductivity at a higher pressure. MgSiO$_3$'s thermal conductivity increases from 6 to 10 W m$^{-1}$ K$^{-1}$ upon compression to 20 GPa[41]. The thermal conductivity of various ferropericlase materials roughly doubles upon compression to 20 GPa[41]. Ice VII's thermal conductivity increases from 4 to 25 W m$^{-1}$ K$^{-1}$ between 2 and 22 GPa[37]. PMMA's thermal conductivity increases by a factor of 3 upon pressurization from 0 to ~10 GPa[42]. The thermal conductivity of muscovite mica, $KAl_2(Si_3Al)O_{10}(OH)_2$, increases by a factor of 10 between 0 and 20 GPa[38]. In a recent review article, Hofmeister reports the $d\Lambda/dP$ for 22 materials[43]. 21 out of 22 materials have positive derivatives that are larger than 3.5% per GPa. Materials whose thermal conductivity do not monotonically increase with pressure often involve a phase transition, *e.g.*, Si[44] or KCl[45]. BAs is not expected to undergo a phase transition below 100 GPa[39].

The unusual pressure independent thermal conductivity of BAs corroborates first-principles predictions that phonon lifetimes in BAs are governed by different processes than in other non-metallic materials. In most materials, three-phonon scattering among two acoustic and one optic mode (*aao*) or three acoustic modes (*aaa*) are the most important type of processes[9,11,46]. However, in BAs, selection rules forbid *aao* processes because the frequency of all optic phonons are more than twice that of the highest frequency of the acoustic mode. As a result, *aaa* and *aaoo* (four-phonon) processes are expected to have the strongest effect on the lifetime of heat-carrying phonons[10]. An *aaoo* process is a four-phonon scattering process that involves two acoustic modes and two optic modes. To understand why BAs has a pressure independent thermal conductivity, we need to consider how *aaa* and *aaoo* processes are affected by pressure.

Upon compression, we expect scattering rates involving *aaa* processes to increase, and the scattering rates for *aaoo* processes to decrease. As mentioned in the introduction, upon compression, there are two major changes to the phonon dispersion of BAs. First, acoustic bunching decreases[10,11], see Figure 1(a). By this, we mean there is a larger difference in frequency between the different acoustic phonon branches. Second, the frequency of optic phonons increases[47].

A decrease in acoustic bunching should increase three-phonon scattering rates by increasing the phase space for *aaa* processes. To understand why, it's instructive to note that an acoustic phonon cannot decay into two acoustic phonons in the same branch[5,19]. This is because it is impossible for three acoustic phonons in the same branch to satisfy crystal momentum and energy selection rules unless the phonon dispersion relation is perfectly linear[5]. Therefore, in the limit that all three acoustic phonon branches were degenerate, the phase space for *aaa* processes would be zero. Of course, transverse, and longitudinal branches are not degenerate in BAs, so *aaa* processes are allowed. But the phase space for such process is more restricted when the frequencies of transverse and longitudinal acoustic branches get closer.

The increase in the frequency of optic phonons should decrease four-phonon scattering rates involving *aaoo* process. Four-phonon scattering rates involving two optic phonons will be proportional to $(1 + n_{o'})n_{o''}$, where $n_{o'}$ and $n_{o''}$ are the thermal occupation factors for the two optic modes. Thermal occupation of optic modes at room temperature will decrease upon compression because frequencies increase.

The above discussion provides a qualitative explanation for why BAs's thermal conductivity depends weakly on pressure. Three-phonon scattering rates increase. Four-phonon scattering rates decrease. These two effects offset each other, and as a result the thermal conductivity remains constant upon compression.

An important goal of our study is to experimentally quantify the relationship between acoustic bunching and three-phonon scattering rates. The weak pressure dependence of BAs thermal conductivity implies the total scattering rate for phonons also depends weakly on pressure. We use this fact to estimate how much three-phonon scattering rates change upon compression to 30 GPa. Raman scattering data and first principles calculation suggest the frequency of optic phonons of BAs increase from ~21 to 25 THz[10,47]. This will lead to a factor of two decrease in thermal occupation of optic modes. Since *aaoo* scattering rates are expected to be proportional to $(1 + n_o)n_o$, we expect the stiffening of optic mode frequencies to decrease *aaoo* scattering rates by a factor of 2. To make crude estimates for how much 3-phonon scattering rates change with pressure, we make two simplifying assumptions. We assume the thermal resistance from three- and four-phonon processes add in series. And, for simplicity, we assume three- and four-phonon scattering processes are of roughly equal importance at 0 GPa. This latter assumption is consistent with first-principles calculations, which predict a thermal conductivity roughly twice what is observed at ambient pressure when four-phonon processes are neglected[8,10,17]. With these assumptions, in order to compensate for a 2× fewer *aaoo* scattering events, *aaa* scattering needs to increase upon compression to 30 GPa by ~50%.

Despite the crudeness of the above analysis, it is in good agreement with first principles calculations. First principles calculations predict that in the absence of four-phonon scattering, Λ of BAs would decrease by 40% upon compression to 30 GPa at 300 K[10]. Additionally, first principles calculations predict *aaoo* scattering rates at frequencies above 5 THz decrease from ~1 GHz to ~0.5 GHz[10].

In Ref. [10], Ravichandran and Broido theoretically studied how temperature and pressure tune the phonon scattering rates in BAs. At 300 K, their calculated thermal conductivity increases by 11% at ~18 GPa and then decreases. They performed calculations on both natural and isotopically pure BAs and claimed that isotope disorder would not affect the pressure dependence. Our measurements on Sample A (1100 W m$^{-1}$ K$^{-1}$) are in excellent agreement with the their calculation results from 0 to 30 GPa at 300 K[10], as shown in Figure 3(b).

Measurements on the two lower-thermal-conductivity BAs samples allow us to evaluate the effect of phonon scattering from crystalline disorder on the Λ(*P*) of BAs. The thermal conductivities of the less-ordered BAs crystals are 40~70% lower than the sample shown in Figure 1(b). In this work, we measured three samples with different defect concentrations to see how defects affect the pressure dependence of Λ$_{BAs}$. It is well known that even a minute concentration of point defects can hugely suppress BAs's thermal conductivity[22–24]. Previous studies suggest there could be many kinds of defects in BAs[22–24,47,48]. A boron or arsenic vacancy concentration of ~1.5×10$^{19}$ cm$^{-3}$ would be sufficient to explain the reduced Λ of our samples[23]. Similarly, As$_B$-B$_{As}$ antisite pair concentration of ~1.5×10$^{19}$ cm$^{-3}$ [48], or carbon impurity concentrations of ~10$^{20}$ cm$^{-3}$ [22], would also explain Λ ~500 W m$^{-1}$ K$^{-1}$ of BAs. To quantitatively evaluate how point-defect disorder affects Λ(*P*) of BAs, we construct a simple relaxation time approximation (RTA) model. The RTA model examines how pressure induced changes in phonon group velocities, phonon-phonon scattering, and defect scattering affect Λ(*P*). The predictions of the RTA are shown in Figure 3(b) as the yellow line. By assuming a pressure independent phonon-phonon scattering rates in BAs, our RTA model agree well with Λ(*P*) of two lower-thermal-conductivity BAs samples. We attribute the pressure independent thermal conductivity of these two samples to pressure independent total phonon-phonon scattering rates. Additional details for the RTA model are in Supplementary Notes.

In conclusion, we measured the pressure dependent thermal conductivity of three BAs samples between 0 and 30 GPa. In contrast to the typical behavior for nonmetallic materials, we observe the thermal conductivity of BAs to be independent of pressure. We attribute this unusual behavior to the pressure independent phonon-phonon scattering rates at P < 30 GPa. We believe the pressure independent scattering rates are caused by a competition between weakening of four-phonon scattering processes and strengthening of three-phonon scattering processes. Our experiments provide the first test of first-principles theories regarding the relationship between phonon dispersion, phonon selection rules, and three- and four-phonon scattering rates[9,10], and improve fundamental understanding of thermal transport in high-thermal-conductivity materials.

**Methods**

**Materials synthesis**

Single crystal BAs (space group: $F\bar{4}3m$) samples are grown by chemical vapor transport (CVT). The reactants are pure boron bulk particles and arsenic lumps. We employ small amount of iodine powder as the transport agent. B and As with a B:As ratio of 1:1.2 along with some iodine were sealed in a fused vacuum quartz tube. The quartz tube was placed in a horizontal two-zone tube furnace with high-temperature zone held at ~890 °C and low-temperature zone held at ~800 °C. Further details about the synthesis can be found in Refs. [49,50].

We used various boron sources in the synthesis processes. Sample A (~1100 W m$^{-1}$ K$^{-1}$) is grown with $^{10}$B isotopes, Sample B (~600 W m$^{-1}$ K$^{-1}$) is grown with $^{nat}$B particles (19.9% $^{10}$B and 80.1% $^{11}$B), and Sample C (~350 W m$^{-1}$ K$^{-1}$) is grown with $^{11}$B isotopes. Samples made of different boron source have different characteristic Raman peaks[50]. We include X-ray diffraction and Raman scattering data on our samples in Supplementary Figure S1.

Detailed defect characterizations, such as transmission electron microscopy (TEM), scanning electron microscopy (SEM), energy-dispersive X-ray spectroscopy (EDX), time-of-flight secondary ion mass spectrometry (TOF-SIMS) etc., can be found in Refs.[12,22,51]. Samples measured in this study were synthesized at the same time with the samples characterized in the mentioned works. TEM shows low dislocation density in high thermal conductivity BAs samples[12]. TEM also shows the presence of mirror twin boundaries. Hall effect measurements indicate our BAs samples are p-type conductive with hole concentrations between $10^{17}$ and $10^{20}$ cm$^{-3}$[22,52]. Impurities such as Si and C are attributed to be the origin of the p-type conductivity. Our previous electron probe microanalysis (EPMA) measurements demonstrate Si impurities of ~0.05 at. % (with a 0.003 at. % detection limit) are present in our BAs samples[22]. Theoretical calculations suggest an impurity concentration of $3.6\times10^{19}$ cm$^{-3}$ for a 500 W m$^{-1}$ K$^{-1}$ BAs sample[22].

**Sample preparation**

We prepared three pieces of BAs for DAC experiments. Two of them (1100 and 600 W m$^{-1}$ K$^{-1}$) were first polished down to $7 \pm 2$ μm. The final thickness was measured with an optical microscope. Then, we deposited a ~ 80-nm-thick Al film on the sample. The other sample (350 W m$^{-1}$ K$^{-1}$) was first coated with an ~90-nm-thick Al film, then being polished from the uncoated side down to $7 \pm 2$ μm. We loaded the samples with 50-80 μm in lateral dimensions into a DAC with a culet size of 300 μm. We loaded ruby spheres alongside the samples as pressure indicators. We used silicone oil (Polydimethylsiloxane, CAS No. 63148-62-9 from ACROS ORGANICS) as the pressure medium for all measurements.

We used 250 μm thick stainless-steel gaskets and pre-indented them in our DAC to a thickness between 30 to 60 μm. Then we drilled holes with a diameter of ~170 μm at the center of the indentations by a laser drill system or an electro-discharge machine. The holes serve as containers for the samples, ruby spheres, and pressure medium.

**Time-domain thermoreflectance (TDTR) in diamond anvil cells**

We measured the thermal conductivity of BAs at ambient and high pressures by TDTR. TDTR is a well-established pump-probe technique. In TDTR measurements, a train of 783-nm-wavelength laser pulses emitted from a mode-locked Ti:sapphire oscillator is split into a pump beam and a probe beam. The pump beam heats the sample at a modulation frequency of ~10 MHz. The probe beam monitors the temperature decay at the sample surface via temperature induced changes in reflectance. The reflected probe beam from the sample surface is collected by a silicon photodiode detector. A lock-in amplifier reads the micro-volt change in voltage output by the detector due to changes in reflected probe beam intensity. The amplifier outputs the in-phase signal $V_{in}$ and out-of-phase signal $V_{out}$ at the ~10 MHz pump modulation frequency. TDTR measurements on the high-purity BAs sample were carried out in University of California Riverside. Further details of our setup can be found in Ref. [53]. TDTR measurements on the other two samples (Sample B and C) were performed at Tsinghua Shenzhen International Graduate School.

Figure 2(a) shows a schematic of the TDTR measurement in a DAC. The pump and probe beams go through the diamond anvil and silicone oil, and reach the sample surface. Figure 2(b) shows a photo of a BAs sample (Sample B) loaded inside a DAC. The pressure of the system is calibrated using the pressure dependent shift of the R1 line in the ruby fluorescence spectrum[54]. We also use the Brillouin frequency of silicone oil as a second measure of pressure[55]. Figure 2(c) shows a Brillouin oscillation that we observe in our experimental TDTR signals. When the pump beam heats the Al surface, it launches a strain wave into the silicone oil medium. The strain wave front moves at the speed of sound of silicone oil. Both the strain wave and Al can reflect the subsequent probe beam. These two reflected probe beams interfere with each other and cause Brillouin oscillations in the $V_{in}$ signal[56].

We used the beam-offset method to measure the laser spot size[57]. The $1/e^2$ radii were 4.5 μm and 5.1 μm for the measurements on Sample B and C (600 and 350 W m$^{-1}$ K$^{-1}$), respectively. For the high-purity sample, we measured the spot size at every pressure, and the $1/e^2$ radii were all around 7 μm.

Prior studies of BAs crystals report a Λ variation of ~10-15% across the crystal surface[14]. We also observed Λ variation on a BAs sample, see Figure 1(b) for the thermal conductivity map. To deal with this concern, we performed TDTR scans at 4-5 locations on the samples at each pressure. However, our results show the variation is only 5-10% at most pressures for all three samples. The thermal conductivity values we report for three samples are the average from the measured spots.

As a control experiment, we measured the pressure dependent thermal conductivity of two MgO samples. The pressure dependence of MgO's thermal conductivity is well studied experimentally[32] and theoretically[10,58]. We prepared the first MgO sample (blue symbols in Figure 4) following similar procedures as Sample C (coat with Al first, then polish to reduce thickness). For the second MgO sample (purple dots in Figure 4), we followed similar procedures as Sample A and B (polish first, then coat with Al). Then we performed TDTR measurements at pressures between 0 and 25 GPa. The 1/e² beam radiuses for measurements of MgO were ~3 and 7 μm for the first and second MgO sample, respectively.

**Data analysis of TDTR under pressure**

We use a bidirectional heat diffusion model to analyze the collected TDTR data[21]. The bidirectional model accounts for heat flow from the Al transducer into both the BAs and silicone oil. The thermal conductivity, heat capacity and thickness of each layer are the input parameters in the heat diffusion model. Therefore, we must estimate how these parameters evolve with pressure to interpret our TDTR data. Below, we describe how we account for the pressure dependence of all parameters.

Prior to loading the sample into the DAC, we measure the Al film thickness by picosecond acoustics[56]. At high pressures, we assume BAs shrinks equally in every direction since BAs is a cubic crystal[39]. If the volume of BAs at pressure $P$ is $V_P$, and the in-plane area is $S_P$, then $S_P = S_0 \cdot (V_P / V_0)^{\frac{2}{3}}$. Here, $V_0$ and $S_0$ are volume and area of BAs at 0 GPa. We assume the in-plane area of Al is equal to $S_P$. Then the thickness of Al at pressure $P$ will be $h_P \approx V_P^{Al}/S_P$. Here, $V_P^{Al}$ is the Al volume at pressure $P$ based on Al's equation of state[59].

To estimate the pressure dependence of Al's heat capacity, we follow Ref. [55], and use a Debye model. For silicone oil, we use previously reported pressure dependent heat capacities and thermal conductivities [60].

To model the pressure dependence of BAs's heat capacity, we use a simple isotropic model for the phonon dispersion. We assume $\omega = v_s k - Ak^2$. Here $\omega$ is the phonon frequency, $v_s$ is the longitudinal or transverse speed, k is the wavevector magnitude, and $A$ is a constant. The value of $A$ is determined by the phonon frequency at the Brillouin zone boundary. We set the values of $v_s$ and $A$ to mimic first principle calculations for phonon dispersion relations vs. pressure[10,61]. Figure 1(a) shows the constructed phonon dispersion relations at 0 and 30 GPa. From the phonon dispersion, we calculate the heat capacities [see equation (1) in Supplementary Materials].

Finally, to interpret the pressure dependent TDTR measurements of MgO, we use the heat-capacity data reported in Ref. [32].

**Thermal conductivity map**

We measured the thermal conductivity map of a BAs single crystal. Our high-purity BAs sample for DAC measurements is cut from this BAs single crystal. We collected the map by a 7-μm laser beam in radius, with step size being 14 μm.

**Spot size dependent TDTR measurements**

Previous works reveal there are obvious spot size effect in the TDTR measurements on BAs[13,14]. We also did TDTR measurements using different laser spot size on Sample A and B at ambient conditions. We used 5×, 10× and 20× objective lenses. The corresponding laser spot sizes were 3, 7 and 16 μm in $1/e^2$ radius for Sample A (1100 W m$^{-1}$ K$^{-1}$), and 3, 6 and 13 μm for Sample B (600 W m$^{-1}$ K$^{-1}$).

## Acknowledgements

This research was supported as part of ULTRA, an Energy Frontier Research Center funded by the U.S. Department of Energy (DOE), Office of Science, Basic Energy Sciences (BES), under Award # DE-SC0021230 (thermal modelling), and by the National Science Foundation (NSF) under Awards # 1847632 and # 1750786 (TDTR measurements). In addition, Zhifeng Ren acknowledges support by the Office of Naval Research (ONR) under MURI Award N00014-16-1-2436 (sample synthesis).

## References

[1]  C. Qian, A. M. Gheitaghy, J. Fan, H. Tang, B. Sun, H. Ye, G. Zhang, *IEEE Access* **2018**, *6*, 12868.
[2]  R. C. Willson, H. S. Hudson, *Nature* **1988**, *332*, 810.
[3]  G. A. Slack, *J. Phys. Chem. Solids* **1973**, *34*, 321.
[4]  S. Tamura, H. J. Maris, *Phys. Rev. B* **1995**, *51*, 2857.
[5]  J. M. Ziman, *Principles of the Theory of Solids*, Cambridge University Press, **1972**.
[6]  C. Kittel, *Introduction to Solid State Physics*, Wiley, Hoboken, NJ, **2005**.
[7]  D. A. Broido, L. Lindsay, T. L. Reinecke, *Phys. Rev. B - Condens. Matter Mater. Phys.* **2013**, *88*, 214303.
[8]  L. Lindsay, D. A. Broido, T. L. Reinecke, *Phys. Rev. Lett.* **2013**, *111*, 025901.
[9]  N. K. Ravichandran, D. Broido, *Phys. Rev. X* **2020**, *10*, 021063.
[10] N. K. Ravichandran, D. Broido, *Nat. Commun.* **2019**, *10*, 827.
[11] L. Lindsay, D. A. Broido, J. Carrete, N. Mingo, T. L. Reinecke, *Phys. Rev. B - Condens. Matter Mater. Phys.* **2015**, *91*, 121202(R).

[12] F. Tian, B. Song, X. Chen, N. K. Ravichandran, Y. Lv, K. Chen, S. Sullivan, J. Kim, Y. Zhou, T.-H. Liu, M. Goni, Z. Ding, J. Sun, G. A. G. Udalamatta Gamage, H. Sun, H. Ziyaee, S. Huyan, L. Deng, J. Zhou, A. J. Schmidt, S. Chen, C.-W. Chu, P. Y. Huang, D. Broido, L. Shi, G. Chen, Z. Ren, *Science* **2018**, *361*, 582.
[13] J. S. Kang, M. Li, H. Wu, H. Nguyen, Y. Hu, *Science* **2018**, *361*, 575.
[14] S. Li, Q. Zheng, Y. Lv, X. Liu, X. Wang, P. Y. Huang, D. G. Cahill, B. Lv, *Science* **2018**, *361*, 579.
[15] J. S. Kang, H. Wu, Y. Hu, *Nano Lett.* **2017**, *17*, 7507.
[16] Q. Zheng, S. Li, C. Li, Y. Lv, X. Liu, P. Y. Huang, D. A. Broido, B. Lv, D. G. Cahill, *Adv. Funct. Mater.* **2018**, *28*, 1805116.
[17] T. Feng, L. Lindsay, X. Ruan, *Phys. Rev. B* **2017**, *96*, 1.
[18] H. Ma, C. Li, S. Tang, J. Yan, A. Alatas, L. Lindsay, B. C. Sales, Z. Tian, *Phys. Rev. B* **2016**, *94*, 220303.
[19] M. Lax, P. Hu, V. Narayanamurti, *Phys. Rev. B* **1981**, *23*, 3095.
[20] M. T. Yin, M. L. Cohen, *Phys. Rev. B* **1982**, *25*, 4317.
[21] D. G. Cahill, *Rev. Sci. Instrum.* **2004**, *75*, 5119.
[22] X. Chen, C. Li, Y. Xu, A. Dolocan, G. Seward, A. Van Roekeghem, F. Tian, J. Xing, S. Guo, N. Ni, Z. Ren, J. Zhou, N. Mingo, D. Broido, L. Shi, *Chem. Mater.* **2021**, *33*, 6974.
[23] N. H. Protik, J. Carrete, N. A. Katcho, N. Mingo, D. Broido, *Phys. Rev. B* **2016**, *94*, 045207.
[24] B. Lv, Y. Lan, X. Wang, Q. Zhang, Y. Hu, A. J. Jacobson, D. Broido, G. Chen, Z. Ren, C.-W. Chu, *Appl. Phys. Lett.* **2015**, *106*, 074105.
[25] P. Jiang, X. Qian, R. Yang, *J. Appl. Phys.* **2018**, *124*, 161103.
[26] Y. K. Koh, S. L. Singer, W. Kim, J. M. O. Zide, H. Lu, D. G. Cahill, A. Majumdar, A. C. Gossard, *J. Appl. Phys.* **2009**, *105*, 054303.
[27] D. G. Cahill, F. Watanabe, *Phys. Rev. B* **2004**, *70*, 235322.
[28] G. T. Hohensee, W.-P. Hsieh, M. D. Losego, D. G. Cahill, *Rev. Sci. Instrum.* **2012**, *83*, 114902.
[29] R. B. Wilson, D. G. Cahill, *Nat. Commun.* **2014**, *5*, 5075.
[30] A. J. Minnich, J. A. Johnson, A. J. Schmidt, K. Esfarjani, M. S. Dresselhaus, K. A. Nelson, G. Chen, *Phys. Rev. Lett.* **2011**, *107*, 095901.
[31] C. Hua, L. Lindsay, X. Chen, A. J. Minnich, *Phys. Rev. B* **2019**, *100*, 085203.
[32] D. A. Dalton, W. P. Hsieh, G. T. Hohensee, D. G. Cahill, A. F. Goncharov, *Sci. Rep.* **2013**, *3*, 2400.
[33] F. Tian, K. Luo, C. Xie, B. Liu, X. Liang, L. Wang, G. A. Gamage, H. Sun, H. Ziyaee, J. Sun, Z. Zhao, B. Xu, G. Gao, X. F. Zhou, Z. Ren, *Appl. Phys. Lett.* **2019**, *114*, 131903.
[34] B. B. Karki, L. Stixrude, S. J. Clark, M. C. Warren, G. J. Ackland, J. Crain, *Am. Mineral.* **1997**, *82*, 51.
[35] B. B. Karki, R. M. Wentzcovitch, S. de Gironcoli, S. Baroni, *Phys. Rev. B* **2000**, *61*, 8793.
[36] S. Daoud, N. Bioud, N. Lebga, *Chin. J. Phys.* **2019**, *57*, 165.
[37] B. Chen, W. P. Hsieh, D. G. Cahill, D. R. Trinkle, J. Li, *Phys. Rev. B - Condens. Matter Mater. Phys.* **2011**, *83*, 132301.
[38] W. P. Hsieh, B. Chen, J. Li, P. Keblinski, D. G. Cahill, *Phys. Rev. B - Condens. Matter Mater. Phys.* **2009**, *80*, 180302(R).
[39] R. G. Greene, H. Luo, A. L. Ruoff, S. S. Trail, F. J. DiSalvo, *Phys. Rev. Lett.* **1994**, *73*, 2476.

[40] R. Boehler, J. Ramakrishnan, *J. Geophys. Res. Solid Earth* **1980**, *85*, 6996.
[41] F. Deschamps, W. P. Hsieh, *Geophys. J. Int.* **2019**, *219*, S115.
[42] W. P. Hsieh, M. D. Losego, P. V. Braun, S. Shenogin, P. Keblinski, D. G. Cahill, *Phys. Rev. B - Condens. Matter Mater. Phys.* **2011**, *83*, 174205.
[43] A. M. Hofmeister, *Proc. Natl. Acad. Sci.* **2007**, *104*, 9192 LP.
[44] G. T. Hohensee, M. R. Fellinger, D. R. Trinkle, D. G. Cahill, *Phys. Rev. B - Condens. Matter Mater. Phys.* **2015**, *91*, 205104.
[45] P. Andersson, *J. Phys. C Solid State Phys.* **1985**, *18*, 3943.
[46] N. K. Ravichandran, D. Broido, *Nat. Commun.* **2021**, *12*, 3473.
[47] X. Meng, A. Singh, R. Juneja, Y. Zhang, F. Tian, Z. Ren, A. K. Singh, L. Shi, J. Lin, Y. Wang, *Adv. Mater.* **2020**, *32*, 2001942.
[48] Q. Zheng, C. A. Polanco, M.-H. Du, L. R. Lindsay, M. Chi, J. Yan, B. C. Sales, *Phys. Rev. Lett.* **2018**, *121*, 105901.
[49] F. Tian, B. Song, B. Lv, J. Sun, S. Huyan, Q. Wu, J. Mao, Y. Ni, Z. Ding, S. Huberman, T. H. Liu, G. Chen, S. Chen, C. W. Chu, Z. Ren, *Appl. Phys. Lett.* **2018**, *112*, 031903.
[50] H. Sun, K. Chen, G. A. Gamage, H. Ziyaee, F. Wang, Y. Wang, V. G. Hadjiev, F. Tian, G. Chen, Z. Ren, *Mater. Today Phys.* **2019**, *11*, 100169.
[51] G. A. Gamage, H. Sun, H. Ziyaee, F. Tian, Z. Ren, *Appl. Phys. Lett.* **2019**, *115*, 092103.
[52] J. L. Lyons, J. B. Varley, E. R. Glaser, J. A. Freitas, J. C. Culbertson, F. Tian, G. A. Gamage, H. Sun, H. Ziyaee, Z. Ren, *Appl. Phys. Lett.* **2018**, *113*, 251902.
[53] M. J. Gomez, K. Liu, J. G. Lee, R. B. Wilson, *Rev. Sci. Instrum.* **2020**, *91*, 023905.
[54] H. K. Mao, J. Xu, P. M. Bell, *J. Geophys. Res.* **1986**, *91*, 4673.
[55] G. T. Hohensee, R. B. Wilson, D. G. Cahill, *Nat. Commun.* **2015**, *6*, 6578.
[56] C. Thomsen, H. J. Maris, J. Tauc, *Thin Solid Films* **1987**, *154*, 217.
[57] J. P. Feser, D. G. Cahill, *Rev. Sci. Instrum.* **2012**, *83*.
[58] X. Tang, J. Dong, *Proc. Natl. Acad. Sci.* **2010**, *107*, 4539 LP.
[59] R. G. Greene, H. Luo, A. L. Ruoff, **1994**, *73*, 11.
[60] W. P. Hsieh, *J. Appl. Phys.* **2015**, *117*, 235901.
[61] S. Daoud, N. Bioud, N. Bouarissa, *Mater. Sci. Semicond. Process.* **2015**, *31*, 124.

**Table of Contents**

Pressure independent thermal conductivity is observed in a high-purity boron arsenide (BAs) single crystal sample between 0 and 30 GPa. Thermal conductivity is measured with time domain thermoreflectance experiments in a diamond anvil cell. The weak pressure dependence of BAs's thermal conductivity is credited to the effect of the phonon dispersion on three- and four-phonon scattering rates.

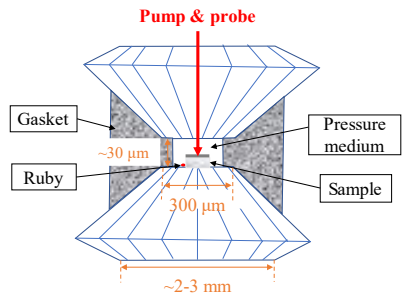